\documentclass[manuscript,screen,sigconf,dvipsnames,anonymous=false,nonacm,authorversion]{acmart}

\usepackage{cleveref}
\usepackage{booktabs}
\usepackage{threeparttable}

\usepackage{pifont}
\newcommand*{\TakeFourierOrnament}[1]{{%
\fontencoding{U}\fontfamily{futs}\selectfont\char#1}}
\newcommand*{\danger}{\TakeFourierOrnament{66}}

\graphicspath{ {./figures/} }

\newcommand*{\fulcioprurl}{\url{https://github.com/znewman01/fulcio/pull/2}}
\newcommand*{\cosignprurl}{\url{https://github.com/znewman01/cosign/pull/118}}

\begin{document}

\title{Reducing Trust in Automated Certificate Authorities via Proofs-of-Authentication}

\author{Zachary Newman}
\email{zjn@chainguard.dev}
\orcid{0000-0001-7238-075X}
\affiliation{%
  \institution{Chainguard, Inc.}
  \country{USA}
}


\begin{abstract}
Automated certificate authorities (CAs) have expanded the reach of public key infrastructure on the web and for software signing.
The certificates that these CAs issue attest to \emph{proof of control} of some digital identity.
Some of these automated CAs issue certificates in response to client authentication using OpenID Connect (OIDC, an extension of OAuth 2.0).
This places these CAs in a position to impersonate any identity.
Mitigations for this risk, like certificate transparency and signature thresholds, have emerged, but these mitigations only detect or raise the difficulty of compromise.
Researchers have proposed alternatives to CAs in this setting, but many of these alternatives would require prohibitive changes to deployed authentication protocols.

In this work, we propose a cryptographic technique for reducing trust in these automated CAs.
When issuing a certificate, the CAs embed a proof of authentication from the subject of the certificate---but without enabling replay attacks.
We explain multiple methods for achieving this with tradeoffs between user privacy, performance, and changes to existing infrastructure.
We implement a proof of concept for a method using Guillou-Quisquater signatures that works out-of-the-box with existing OIDC deployments for the open-source Sigstore CA, finding that minimal modifications are required.
\end{abstract}

\maketitle

\section{Introduction}
Let's Encrypt~\cite{aas2019let}, a free and automated certificate authority (CA), has dramatically increased TLS adoption on the web.
Inspired by this, developers have created new automated CAs for managing code signing certificates and SSH credentials.
The ACME protocol that Let's Encrypt uses was standardized in RFC 8555~\cite{rfc8555}, and the IETF's ACME working group has produced additional standards-track and draft documents for a variety of extensions.

Automated CAs can lower barriers to adoption of security technologies, like encrypted web traffic and digital signatures for software.
This is especially critical for the security of open source software, whose maintainers often will not buy or cannot afford code signing certificates from established CAs, which can cost hundreds of US dollars.
While PGP~\cite{garfinkel1995pgp} (and the open-source GPG implementation) keys are free, managing them can be challenging for even sophisticated users~\cite{whitten1999johnny,ruoti2015johnny}.
Automated certificates eliminate the need for managing long-lived key pairs.
Further, if obtaining a certificate is sufficiently fast and easy, these certificates can be extremely-short lived (on the order of minutes), reducing the risk of compromise for a long-lived key.

However, an increased reliance on automated CAs comes with risks: if these CAs are compromised, they stand in a position to impersonate large numbers of users.
Here, we propose a method for reducing the trust required by clients in automated certificate authorities: proofs-of-authentication.
If CAs can prove that they issued a certificate in response to an authentication event, clients can verify this proof.
A malicious CA could not successfully issue a certificate that clients will accept without a corresponding authentication token.
In this work, we focus on automated CAs based on OpenID Connect (OIDC)~\cite{oidc}, contributing a method based on the Guillou-Quisquater~\cite{guillou1988practical} identification scheme that allows CAs to prove that they have seen a corresponding OIDC token.

\section{Background}
Here, we discuss relevant background for this system.

\paragraph{Automated Certificate Authorities}
Public key infrastructure (PKI) typically includes \emph{certificate authorities} (CAs), which issue \emph{certificates} with \emph{subjects} that describe identities.
Some PKI designs require expensive, manual registration procedures,
The archetypal example is Extended Validation~\cite{cab-forum-ev} (EV), which claims to guarantee that a certificate belongs to a specific legal entity.
Despite the expense, EV certificates do not, in practice, provide additional security~\cite{biddle2009browser,jackson2007evaluation,maimon2020extended}.
Instead, recent years have seen a number of automated CAs.

Automated CAs issue certificates in response to a proof of control of a digital identity.
For instance, in web PKI, Let's Encrypt~\cite{aas2019let} introduced the Automated Certificate Management Environment (ACME) protocol~\cite{mccarney2017tour,rfc8555}, issuing certificates to any party that proves control of a domain name in DNS.
Automated CAs have a number of advantages: namely, they are easy-to-use, free, and fast.

\paragraph{Certificate formats.}
A CA issues certificates in a particular format; participants in the PKI must agree on this format out-of-band.
In this work, we assume without loss of generality that CAs use the X.509~\cite{x509} format (which is used in web PKI).
X.509 certificates support extensions: embedding additional authenticated arbitrary fields in the signed certificate.
Extensions are identified with Object Identifiers (OIDs).

\paragraph{OIDC} OpenID Connect (OIDC)~\cite{oidc} is an extension to OAuth 2.0~\cite{rfc6749} for authentication.
In OIDC, an identity provider issues signed bearer tokens to clients, which can present them to downstream audiences to authenticate.
These tokens are in the JSON Web Token (JWT)~\cite{rfc7519} format and comprise a header, body, and signature, each base64-encoded and separated by periods.
These audiences check the signatures on the tokens against the identity provider's verification key in the JSON Web Key (JWK)~\cite{rfc7517} format, which they fetch from a configured URL.
These verifications keys rotate frequently.
Popular OIDC identity providers include web services (Microsoft, Google, Facebook, GitHub, GitLab), identity management companies (Okta, Auth0), and nonprofits (Mozilla).
OIDC tokens can correspond to human identities (usernames or emails) or machine identities (specific workloads on a cloud provider).

OIDC can be used as the basis for an automated CA.
Step Certificates~\cite{step-ca} uses OIDC to automatically issue SSH certificates, and Sigstore~\cite{newman2022sigstore} uses OIDC to issue X.509 code signing certificates.

In theory, OIDC tokens themselves could be used as certificates.
The proposed Demonstrating Proof of Possesssion at the Application Layer (DPoP) extension to OAuth~\cite{ietf-oauth-dpop-16} binds OAuth tokens to public keys, but requires invasive changes to OIDC providers and relying parties.
OpenPubkey~\cite{heilman2023openpubkey} implements the same functionality but is compatible with the current OAuth specification.
However, both solutions require exposing OIDC tokens, which were originally intended to be kept secret, publicly.
While this should not be an issue for compliant relying parties, which should reject tokens with an unknown \emph{audience}, noncompliant implementations might unwittingly continue to treat these documents as bearer tokens, rather than certificates.

\paragraph{CA risk}
CAs play a trusted role in a PKI.
A compromised or malicious certificate authority could issue certificates with arbitrary subject to impersonate any identity.
There are several techniques to mitigate this risk.
First, users should trust only trustworthy CAs: organizations like the CA/Browser Forum~\cite{cab-forum} (CAB Forum) establish and enforce rules for CAs.
Second, \emph{certificate transparency}~\cite{laurie2014certificate,rfc6962,rfc9162} makes all of the certificates issued by a CA public.
This enables the CA itself to detect compromise, or for identity holders to notice certificates they did not request; after a compromise, investigators can identify \emph{all} maliciously issued certificates.
Finally, users can require signatures from multiple CAs, reducing the impact of the compromise of a single CA~\cite{braun2014trust}.
However, none of these mitigations can entirely prevent attacks from malicious CAs.

\paragraph{Ledgers and Certificate Transparency}
Transparency logs~\cite{laurie2014certificate} are public logs of some structured content.
These logs are typically backed by a tamper-proof authenticated data structure.
The logs have a succinct digest, which clients can store.
The log operator can provide proofs of membership with respect to the log's digest for entries in the log.
The log operator can also provide consistency proofs to support client updates: proofs that one digest represents a log which is a prefix of the log represented by a second digest.

While a clients can verify the consistency of a transparency log \emph{after} they first fetch a digest, the log may \emph{equivocate} and serve different digests to different users.
To prevent this, some such systems rely on a consistency protocol like that of \citet{parakeet} in Parakeet.
In such protocols, \emph{witnesses} sign ``valid'' log digests and publish these signatures via the log itself.
Clients check for a quorum of witness signatures before accepting a digest.
The definition of ``valid'' may change between systems.
Witnesses may only serve to prevent equivocation, or they may check the consistency of the log, or they may check the correctness of the log (either entry-by-entry, or by treating the log contents as transitions in a state machine).

Certificate transparency (CT)~\cite{rfc6962,rfc9162} is a system of transparency logs of certificates issued in an X.509-based PKI; the most prominent deployment is in web PKI.
In CT, a CA creates a \emph{precertificate} (which are not valid certificates) which it sends to CT logs.
The CT logs create Signed Certificate Timestamps (SCTs), indicating their intent to merge the certificate into the log.
The CA embeds the SCT into the final certificate.
There is no prescribed consistency protocol.
There \emph{are} auditors, who check for log consistency (that no entries are removed).

\section{Setting and goals}
\label{sec:setting-and-goals}
We consider a public key infrastructure including the following parties with the given roles:

\begin{itemize}
\item \textbf{Requester}: wants a certificate linked to a given identity.
\item \textbf{Identity provider (IdP)}: authenticates the requester's control of that identity.
\item \textbf{Certificate authority (CA)}: issues certificates.
\item \textbf{Certificate Transparency Log (CTL)}: logs issued certificates.
\item \textbf{Verifier}: verifies the certificates.
\end{itemize}

\begin{figure}[t]
  \includegraphics[width=\linewidth]{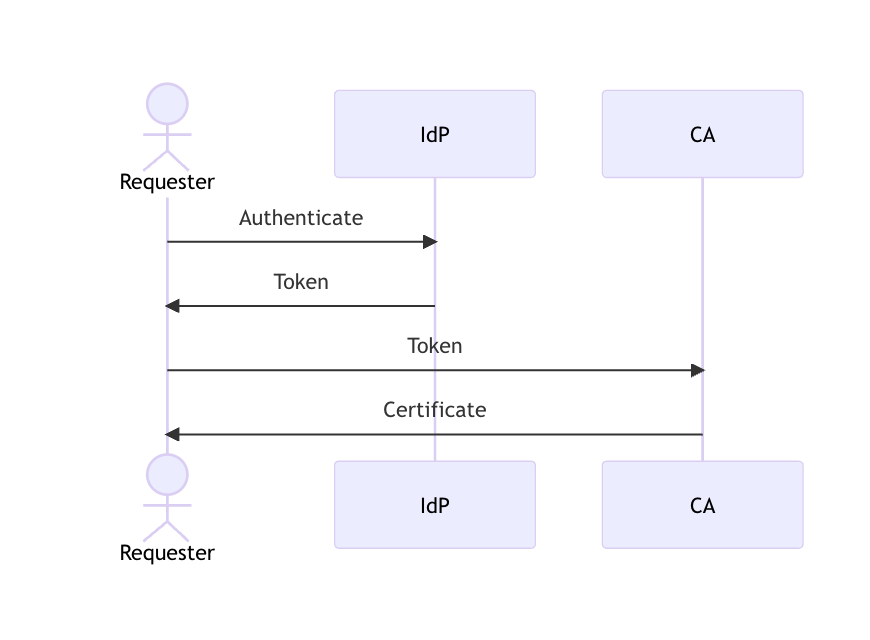}
  \caption{Certificate request flow, including the requester, identity provider (IdP) and certificate authority (CA).}
  \label{fig:auth-flow}
\end{figure}

\noindent
To get a certificate, the requester generates a key pair, authenticates with the IdP to receive and identity token and presents the identity token along with the public verification key to the CA, which verifies the identity token for the IdP and issues a certificate (\cref{fig:auth-flow}).
The requester can then sign using the signing key.
The verifier checks the certificate itself against the CA's root certificate, then uses the certificate to check signatures from the requester.
Importantly, the certificate should be verifiable well after the lifetime of the OIDC token (this allows converting momentary authentication into long-lived signatures).
For this work, we assume that the IdP issues OpenID Connect (OIDC) tokens~\cite{oidc} and that the certificate authority issue X.509 certificates~\cite{x509}.
Additionally, we introduce a new party, the \textbf{JWK Ledger}, which operates a signed, public, tamper-proof log of IdP verification keys.

Our objective in this work is to prevent the CA from improperly issuing certificates---namely, from issuing a certificate without a requester provindg a valid and signed identity token.
Otherwise, we make the typical security assumptions of automated signing: we assume that the verification keys for the CA, JWK Ledger, and CTL are distributed securely out-of-band, and that all parties can securely fetch the verification keys for the identity provider (which follows from the security of web PKI).
We assume for the sake of presentation that the CA only supports one IdP, and that this identity provider is known to all participants; it is straightforward to extend the protocol to support multiple IdPs.
We make the same cryptographic assumptions as in the baseline automated signing protocol (namely, the security of a signing algorithm).
We make an additional cryptographic assumption: that the algorithm used to sign the OIDC identity token must admit a proof of knowledge of signature (which will be defined below).
All compliant OIDC IdPs must support RSA SHA-256; we can construct a proof of knowledge of signature for this scheme from the RSA assumption (which we have already made in using RSA SHA-256); this proof of knowledge can be made more efficient by a constant factor by under the adaptive root assumption~\cite{wesolowski2019efficient}.
Our system must meet the following goals:

\begin{itemize}
  \item \textbf{Completeness:} given a valid authentication token, a CA following the protocol should produce a certificate that the verifier accepts.
  
  \item \textbf{Unforgeability:} no CA should be able to create a valid certificate without seeing a corresponding valid and signed identity token.

  \item \textbf{Deployability:} the system should require no changes to the identity providers and be compatible with currently-deployed OIDC features.
  Changes to clients, CAs, and verifiers are permitted, along with the addition of the JWK Ledger.

  \item \textbf{Performance:} the costs to create a certificate and to verify a certificate must be negligible compared with existing certificate creation costs; the size of the modified certificates must be no more than double the size of the current certificates.

  \item \textbf{Replay protection:} no \emph{new} party should be able to authenticate using the requester's identity (the requester and identity provider trivially can, and the CA might be able to perform a replay attack).
\end{itemize}

\noindent
We give more precise definitions for completeness, unforgeability, and replay protection in \cref{sec:defs-formal}.

\section{Design}
\label{sec:design}
In the case of automated CAs, we can provide better security, at least in theory: certificates are issued only after a client successfully completes an authentication process.
If signers use the authentication challenge directly as a form of certificate, verifiers no longer need to trust a CA.
However, this is circular: if we had an authentication system in which the identity provider acted as a certificate authority, we would no longer need an additional certificate authority.
The primary challenge here is engineering effort: adding CA capabilities to every identity provider is simple in theory---identity providers just issue certificates where they had been giving out bearer tokens---but in practice, convergence on a standard would take several years, and deployment would take several more.

Instead, we propose the use of existing third-party certificate authorities to turn existing authentication protocols into verifiable certificates.
There are several challenges in doing this, both in general and for OIDC in particular:

\begin{enumerate}
\item \textbf{Interactivity.} Many authentication protocols are interactive; a certificate must be verifiable non-interactively.
\item \textbf{Replay attacks.} Embedding a bearer token into the certificate would enable non-interactive verification, but also enable anybody seeing the certificate to impersonate the user.
\item \textbf{State.} OIDC tokens are signed by keys which rotate frequently (daily is recommended), so verifying tokens later may not be possible.
\end{enumerate}


\noindent
In our proposed design, the CA issues certificates automatically when presented with an OIDC token, but now the CA embeds a \emph{proof of authentication} into the certificate.
Additionally, the CA records the evidence used in authentication (namely, the public keys used to verify the tokens) in a public JWK Ledger.
Together, this allows verifiers to check the tokens directly.

\subsection{Proofs of authentication}
A proof of authentication lets the CA prove to verifiers that they in fact did see a valid authentication token for the given subject.
This prevents CAs from issuing certificates that have not been requested.

\paragraph{Insecure strawman}
The simplest proof of authentication would embed the full token directly into the certificate, and let verifiers check the signature on the token directly.
The primary problem with this mechanism is the potential for replay attacks: a verifier could take this token, which is a bearer token, and present it to themselves authenticate.
While OIDC contains multiple mechanisms to combat these attacks---scoped audiences and expiration times---bearer tokens are meant to be a secret:  verification implementations might omit these checks, and even if properly implemented, a verifier could quickly (before token expiry) send the token back to the CA (same audience) and successfully authenticate.

\paragraph{General purpose zero-knowledge proofs}
Instead, we can have the CA demonstrate proof of knowledge of a valid authentication token.
One option is to use a general-purpose zero knowledge proof of knowledge to prove a statement of the rough form:
\begin{quotation}
For public OIDC verification key $K$, identity $I$, and time $T$, I know bytes $X$ such that $X$ parses as a valid OIDC token that has been signed by $K$ \emph{and} $X$ has the subject $I$ \emph{and} $X$ was valid at time $T$.
\end{quotation}
\noindent
Then, verifiers check that $K$ was a valid verification key for the appropriate provider, that the times match, and so on.
However, this is expensive and error-prone, requiring the implementation of the full OIDC verification procedure inside of the zero-knowledge proof.

\paragraph{Proof-of-knowledge of signatures}
By relaxing the requirement that this proof of knowledge be zero-knowledge, we can achieve much better performance.
An OIDC token consists of three parts: a header (which identifies the signing key and algorithm used), a body (which includes the subject of the token, its validity period, and other data), and a signature over the body.
Without the signature, the token cannot be used to authenticate.
Therefore, the CA can embed the header and the body into the issued certificate.
Instead of a signature, they embed a proof of knowledge of a valid signature~\cite{nguyen99zkpop} for the given message and given public key.

The OIDC specification requires that providers implement RSA signatures for OpenID tokens; RSA signing admits proofs of knowledge of signatures using the Guillou-Quisquater (GQ) identification scheme~\cite{guillou1988practical,heilmanslack}.
Because this scheme achieves only knowledge soundness of $\frac{1}{e}$, where $e$ is the public RSA exponent~\cite{bangerter2005efficient}, the scheme would need to be repeated $\lambda / \lg(e)$ times to achieve $\lambda$-bit security.
\citet{boneh2019batching} show that introduce a short, trapdoor-free common reference string allows achieving the needed soundness with only one repetition under the adaptive root assumption~\cite{wesolowski2019efficient}.

\paragraph{Self-signed certificates}
Given that this work reduces the trust in CAs because users can verify the check that the CA does, can we totally eliminate the role of the CA?
Trust in the CA is required because otherwise a user could extract a proof-of-authentication from a certificate and embed it into their own self-signed certificate: it's important that a trusted party CA create the proof-of-authentication.

\subsection{Certificate issuance}
Here, we describe the certificate issuance procedure for an OIDC-based automated CA with proof of authentication.
During certificate issuance, such a CA must map the various claims in an OIDC token to X.509 extensions---for instance, the \texttt{sub} claim might map to the \texttt{SubjectAlternativeName} extension.
The exact method will vary, but should be deterministic and known to all parties.
We also define additional extensions to X.509 for proof of authentication.
For the sake of example, we use the \texttt{1.3.99XX} prefix, which is reserved for private use.
Real deployments should register proper extensions.
On receipt of a certificate issuance request, which includes an OIDC token and a public key, the certificate authority will:

\begin{enumerate}
\item Validate the authentication token per the OIDC specification~\cite{oidc} (\S 3.1.3.7 \S 3.2.2.11, or \S 3.3.3.7, depending on the flow used; encrypted tokens are disallowed), including fetching the verification keys of the IdP. If this fails, abort.
\item If the set of the IdP's verification keys has changed, send the new set to the JWK Ledger.
\item Populate an X.509 precertificate based on the OIDC token. 
\item Write the JWT header and body (still base64-encoded) as bytes in the \texttt{1.3.9901} extension of the precertificate.
\item Compute a proof of knowledge of signature for the JWT signature and the IdP's verification key. Encode this proof of knowledge of signature as bytes in the \texttt{1.3.9902} extension of the precertificate.
\item Send the precertificate to the CTL, and wait for a certificate with an embedded Signed Certificate Timestamp (SCT).
\item Send the certificate to the requester.
\end{enumerate}

\noindent
The SCT includes a signature over the certificate body and a timestamp.

\subsection{JWK Ledger}
Verifying OIDC identity tokens requires the public key of the identity provider.
However, OIDC was intended for moment-in-time authentication, and these public keys are not available after they are rotated.
The verifier cannot simply get the requisite public key from the CA, as a malicious CA could provide a bad public key and forge tokens signed by this key.

To enable verification for OIDC tokens after the fact, the CA records public keys in a ``JWK Ledger.''
Specifically, the CA already maintains a local cache of the set of verification keys for an identity provider, fetched from a known per-provider URL, in order to perform OIDC verification. 
Whenever a key is added or removed from this key set, the CA sends the current state of the cache to the ledger.
The ledger independently verifies the key set and records all changes to it (along with a timestamp).
Users query the ledger for the key set for an IdP at a particular time.

To decrease the risk that the JWK Ledger will present false information to users, this ledger is backed by a transparency log.
The pace of updates to this log should be relatively low, occurring only when the IdP rotates verification keys.
Therefore, witnesses for the log can verify the current state of the key set on each update; they also check that no entries other than the given key set change have been added to the log.
Clients can requires a quorum of witnesses on the JWK Ledger digest.
When a client requests the key set for a given timestamp, the ledger serves two entries.
The client checks that the timestamp of the first entry precedes the requested timestamp, that the timestamp of the second entry follows the requested timestamp, and that the entries are adjacent in the log.
This convinces the client that the key set was valid at the given time.

\subsection{Verification}
To verify a certificate, the verifier will:

\begin{enumerate}
\item Verify the certificate as before (including SCT verification).
\item Extract the timestamp from the SCT. In later steps, the verifier will use this timestamp as the ``current time.''
\item Extract the JWT header and body from the X.509 extensions in the certificate.
\item Validate the authentication token per the OIDC specification~\cite{oidc} (\S 3.1.3.7 \S 3.2.2.11, or \S 3.3.3.7, \emph{except} that the verifier should not check the signature or nonce, and the verifier should check that the token's \texttt{aud} claim identifies the CA.
\item Fetch and verify the set of the IdP's verification keys at the ``current time'' from the JWK Ledger, as above.
\item Extract the proof of knowledge of signature from the certificate. Check the proof against the resolved set of verification keys.
\item Check that the certificate was populated correctly based on the body of the OIDC token, using the same (public) rules as in certificate issuance.
\end{enumerate}

\section{Security analysis}
\label{sec:security-analysis}

Per \cref{sec:setting-and-goals}, we have two security goals: unforgeability and replay protection.
Here, we give intuitive definitions and security arguments for the system instantiated with the Guillou-Quisquater (GQ) proof-of-knowledge-of-signature; see \cref{sec:defs-formal} for precise definitions and arguments.

\paragraph{Unforgeability.}
Unforgeability requires that an attacker controlling a CA cannot create certificates that a verifier accepts for a given subject without seeing a valid OIDC token for that subject.
This property follows from the proof-of-knowledge property of our proof of authentication: an attacker who could forge such certificates would be able to create a valid, signed OIDC token against a randomly-generated OIDC identity provider public key, contradicting the security of the underlying RSA signature scheme.

\paragraph{Replay protection.}
Replay protection requires that an attacker cannot product a valid OIDC token, even after seeing the certificate in this scheme.
The naive system that embeds the signed OIDC token into the certificate provides unforgeability but not replay protection.
Replay protection in this system follows from the zero knowledge property of the GQ identification scheme: an adversary capable of forging such tokens must be able to forge valid GQ signatures without seeing the underlying RSA signatures.

\newcommand{\good}{{\textcolor{ForestGreen}{\ding{51}}}}
\newcommand{\warn}{{\textcolor{BurntOrange}{\danger}}}
\newcommand{\bad}{{\textcolor{Maroon}{\ding{55}}}}

\begin{table*}
  \begin{minipage}{\linewidth}
  \centering
  \begin{threeparttable}
  \begin{tabular}{@{}l@{}ccc@{}}
  \toprule
  \textbf{Attacker controls...}         & \multicolumn{3}{c}{\textbf{Protected from...}}\\
                                        & Forging certs (current) & Forging certs (historical) & Replaying OIDC tokens \\
  \cmidrule{2-4}
  OIDC Identity Provider                & \bad                    & \bad                       & \bad  \\
  Certificate Authority (CA)            & \good                   & \good                      & \warn  \\
  JWK Ledger                            & \good                   & \good                      & \good \\
  Witnesses                             & \good                   & \good                      & \good \\
  CT Log                                & \good                   & \good                      & \good \\
  CA, JWK Ledger, \& Witnesses          & \bad\rlap{\tnote{a}}    & \good                      & \warn \\
  CA, JWK Ledger, Witnesses, \& CT Log  & \bad\rlap{\tnote{a}}    & \bad\rlap{\tnote{a}}       & \warn \\
  \bottomrule
  \end{tabular}
  \caption{An attacker controlling various subsets of parties}
  \label{tbl:compromise}
  \begin{tablenotes}
  \item[a] Can be detected by third parties.
  \end{tablenotes}
  \end{threeparttable}
  \end{minipage}
\end{table*}

\bigbreak
\noindent
The above properties assume, for simplicity of definition, that OIDC public keys are fixed.
Provided that the JWK Ledger and witnesses behave honestly, the same properties hold in a setting with changing OIDC public keys.
\Cref{tbl:compromise} shows what happens if these (and other) parties are compromised.
As expected, OIDC identity provider compromise is fatal: the identity provider is the source of truth for identities, and can consequently authenticate as any identity and request certificates in their name.
A compromise of the CA permits OIDC token replay: they can reuse the token.
The \texttt{aud} feature and expiry times of JWTs mitigate this risk, since conformant relying parties should reject tokens with the \texttt{aud} set to the CA.
Otherwise, compromise of any other individual party does not give the attacker the ability to forge certificates or replay OIDC tokens. 
However, compromise of \emph{all} parties besides the identity provider allows forging certificates, both historically and in the present.
Still, always-online clients who have previous state from interaction with this system can detect that the JWK Ledger and CT log have been modified.

The above analysis uses a relatively simple model of OIDC.
The OIDC specification contains many features for mitigating specific compromises that are outside the simple threat model for OIDC but may be relevant in practice.
Once such feature is the \emph{nonce}.
OIDC supports three ``flows:'' the code flow, the implicit flow, and the hybrid flow.
In both the hybrid and implicit flow, OIDC verifiers must check a nonce that they provided earlier.
CAs using OIDC can perform this check, but the end-user verifiers cannot, as they do not know the nonce.
This creates a very minor risk: a malicious CA who obtains a replayed OIDC token that would have been detectable via the nonce can now forge a valid certificate embedding this token that end-user verifiers will accept.
Otherwise, end user verification of the OIDC token is identical to verification by an OIDC relying party, except for the use of a GQ signature in place of an RSA signature (which is analyzed above).

\paragraph{Privacy tradeoffs.}
This system has an obvious privacy downside: it leaks all of the \emph{body} of the OIDC token. 
While this isn't enough to impersonate the subject of the token, this information could be sensitive.
Taking the \texttt{accounts.google.com} JWT tokens as an example~\cite{google-oauth}, these tokens embed only a limited amount of relevant infomration: the email of the user, whether the user's email has been verified, and the domain of the user's Google Cloud organization (which is likely the same as the domain from the user's email).
These are likely already embedded in the ultimately issued certificate.
When explicitly requested, the token includes a name, locale for the name's encoding, profile URL, and profile picture URL; this should not be the case for the automated CA use case.
The remaining claims are not tied to the specific user, and represent things like the token's expiry time.
While this information is limited, in settings where even email addresses should be hidden from the certificate~\cite{speranza}, embedding OIDC tokens could pose privacy and compliance challenges.
In this case, the general-purpose zero-knowledge proof construction would address these issues, though it would come with corresponding performance degradation.

\paragraph{Proof-of-authentication replay}
The above definition of replay protection only protects against replay of the signed OIDC token, not replay of the proof of authentication.
If two independent CAs implemented the same scheme, a malicious CA could copy the proofs of authentication.
However, we require that verifiers check the audience of the JWT when verifying the proof of authentication.
This would detect and reject such an attack.

\section{Proof-of-concept implementation}
We created a prototype implementation of the Guillou-Quisquater proof-of-authentication scheme for Sigstore~\cite{newman2022sigstore}, an automated OIDC PKI.
This involves modifications to Fulcio, the Sigstore CA\footnote{\fulcioprurl} and Cosign, the Sigstore client~\footnote{\cosignprurl}.
The change to each requires about 500 lines of Go code, including whitespace.
Most of this (over 70\%) is the implementation of the underlying cryptographic primitives for the non-interactive variant of the GQ identification scheme.
This implementation could be factored out into a library and shared across the client and CA, as well as any other potential users of the GQ scheme.
The remainder is changes to the CA itself, to embed the JWT header and body (but not the signature) and the GQ proofs into the certificate, and to the client, to check the GQ proofs.

\paragraph{Deployment Considerations}
In theory, OIDC is standardized enough that an OIDC-based automated CA can accept new identity providers without modification; in practice, many providers create slightly nonstandard tokens.
For this reason, Sigstore uses a federated OIDC provider, Dex~\cite{dex}, effectively as a proxy to smooth out differences between upstream identity providers.
This deployment of Dex is conceptually part of Fulcio, the Sigstore CA, from a trust perspective.
This means that Fulcio never actually sees a JWT from many of the configured identity providers.
For our prototype implementation, we provided GQ proofs for the Dex-issued tokens; a real implementation must modify Sigstore to create these proofs for the upstream tokens.

\section{Discussion and conclusion}
This work proposes the use of proof-of-authentication for improving the verifiability of automatically-issued signing certificates, mitigating the fallout of certificate authority (CA) compromise.
It focuses on OIDC-based automated CAs.

\subsection{Related work}

\paragraph{Demonstrating Proof-of-Possession at the Application Layer (DPoP)}
In DPoP~\cite{ietf-oauth-dpop-16}, a proposed extension to OAuth 2.0, end users create asymmetric signing keypairs, sending the public verification key to the identity provider during authentication.
Then, the identity provider signs a JWT (a DPoP-bound access token) that includes this verification key as a claim.
This makes the access token, in effect, a certificate.
Clients sending requests authenticated with a DPoP-bound access token must use the corresponding signing key to create a DPoP proof JWT for each request, which is a signature over the HTTP URI (and method) of the request, though not the request contents.
By requiring proof-of-possession of the signing key associated with the JWT, we prevent an attacker who has obtained a JWT from replaying it.

However, though these access tokens function in some sense as certificates, the goal of the DPoP specification is to provide additional assurance when using JWTs as bearer tokens with requests to prevent replay specifically: DPoP-bound access tokens should not be made public.
Further, DPoP-bound access tokens can only be verified until the identity provider rotates their keys.
Finally, DPoP is still a draft, and not implemented by today's identity providers; deploying DPoP requires modifications to identity providers, which violates the ``deployability'' goal of this work (\cref{sec:setting-and-goals}).

Once deployed, a CA could require DPoP proofs during certificate issuance.
In a system with proofs-of-authentication, clients could then verify DPoP proofs along with the DPoP access token, partially mitigating the ``minor risk'' caused because clients cannot verify nonces (see \cref{sec:security-analysis}).
This is, however, not a replacement for proofs-of-authentication.

\paragraph{Other OAuth extensions.}
RFC 8705~\cite{rfc8705} introduces a way to bind JWTs to a mutual TLS (mTLS) certificate held by a client.
This solves a similar problem to DPoPs (tighter binding of requests to).
A draft proposing OAuth 2.0 token binding \cite{ietf-oauth-token-binding} does much the same, using raw public keys rather than certificates.
However, the mTLS certificate and corresponding key pair should only be used for TLS authentication, not code signing; a similar argument against using the token-binding key pair for code signing applies.
These methods also have the limitation of only allowing verification until the identity provider's OIDC public keys rotate.
Something like the JWK Ledger would be required to allow historical verification.
While a separate proposal could extend OAuth tokens to address these issues, it has the disadvantage of requiring identity provider changes, limiting its ability to be immediately deployed.

\paragraph{OpenPubkey}
OpenPubkey~\cite{heilman2023openpubkey} proposes a clever trick for using the \texttt{nonce} of a JWT to embed a commitment to a public key.
This allows using existing identity providers to issue ``JWT certificates'' without modification.
However, this method is nonstandard and has a backwards compatibility issue: OIDC verifiers which predate (or have not adapted to) OpenPubkey will accept ostensibly ``bound'' tokens.
This means that OpenPubkey tokens should not be made public unless the audience is known to support it.
However, this is sufficient for a use case similar to that of DPoP (replay protection), with OpenPubkey providing only additional support.
Further, OpenPubkey also has no notion of historical public keys for OIDC issuers.
  
\paragraph{Verifiable credentials}
Verifiable credentials~\cite{verifiable-credentials} (VCs) are digital credentials created by an issuer and held by a holder.
While the core protocol does not specify a proof mechanism, VCs support use with proof mechanisms like JSON-LD proofs or anonymous credentials~\cite{chaum1983blind,camenisch2001efficient}. 

These could, in theory, be used to create long-lived signatures bound to the credentials as well.
In this setting, the verifiable credential issuer is, in effect, already functioning as a certificate authority, so there is no need for an additional CA.

\subsection{Future work}
Future work might extend the range of available proofs of authentication.
OIDC mandates RSA signatures, but allows any RFC 7518~\cite{rfc7518} signatures, including ECDSA.
Further, additional future signature algorithms could be added.
Discrete log-based signature schemes, like ECDSA, support proofs-of-knowledge-of-signatures~\cite{nguyen99zkpop} and might be good candidates.

Additionally, other automated CAs could be made verifiable using similar techniques.
ACME~\cite{mccarney2017tour} uses reachability via DNS or other protocols, and therefore may not be a good fit, but extensions to ACME like RFC 8823~\cite{rfc8823}, which uses email-based signatures, could embed DKIM~\cite{rfc6376} metadata into the ultimately-issued certificate.

While for immediate deployability, a primary design goal was avoiding the requirement to modify existing authentication schemes; a longer-term effort could use authentication schemes designed to support proof-of-authentication natively.

Finally, we invite more investigation into the JWK Ledger.
Would techniques from key transparency apply?
Could the set of keys for a given provider be encoded more efficiently, perhaps using an interval tree data structure?

\begin{acks}
We would like to thank Ethan Heilman, who found the Guillou-Quisqater construction and for helpful comments and throughout.
Further, we thank Hayden Blauzvern and Santiago Torres-Arias for many discussions on Sigstore CA security.
We also thank the Sigstore authors, maintainers, contributors, and community.


\end{acks}

\bibliographystyle{ACM-Reference-Format}
\bibliography{sample-base}

\appendix
\section{Security Definitions and Arguments}
\label{sec:defs-formal}

Here, we give definitions of the protocols used and brief definitions of security and arguments that these properties hold.
Throughout, fix security parameter $\lambda$.

\paragraph{OpenID Connect.}
We use a simple model of the OpenID Connect protocol with three randomized algorithms:

\newcommand{\alg}[1]{\mathsf{#1}}
\newcommand{\pk}{\mathsf{pk}}
\newcommand{\sk}{\mathsf{sk}}
\newcommand{\pkca}{\pk_\mathsf{ca}}
\newcommand{\skca}{\sk_\mathsf{ca}}
\newcommand{\pkoidc}{\pk_\mathsf{idp}}
\newcommand{\skoidc}{\sk_\mathsf{idp}}
\newcommand{\id}{\mathsf{id}}
\newcommand{\tok}{\mathsf{tok}}
\newcommand{\cert}{\mathsf{cert}}
\newcommand{\yes}{\mathtt{yes}}
\newcommand{\no}{\mathtt{no}}
\newcommand{\adv}{\mathcal{A}}
\newcommand{\negl}{\mathsf{negl}}

\begin{itemize}
  \item $\alg{OIDC.Generate}(\lambda) \to (\skoidc, \pkoidc)$: generate a signing and verification key for an OIDC identity provider.
  \item $\alg{OIDC.Issue}(\skoidc, \id) \to \tok$: issue a token for the given identity.
  \item $\alg{OIDC.Verify}(\pkoidc, \mathsf{tok}, \id) \to \yes/\no$: validate a token for the given identity against a known verification key.
\end{itemize}

\noindent
Authentication itself is orthogonal to this work. These algorithms must satisfy correctness (verifying an honestly-issued token must succeed) and unforgeability (the probability that a computationally bound attacker can issue a valid token for an identity of their choosing that verifies successfully against a correctly-generate public key is negligible).

\paragraph{Automated OIDC Certificate Authorities.}
Here, we give a model of an automated OIDC certificate authority supporting embedded proofs-of-authentication.
For simplicity, we ignore rotation of OIDC verification keys and the JWK Ledger; these are analyzed in \cref{sec:security-analysis}.

\begin{itemize}
  \item $\alg{CA.Generate}(\lambda) \to (\skca, \pkca)$: generate a signing and verification key for a certificate authority.
  \item $\alg{CA.Issue}(\skca, \pkoidc, \tok) \to \cert$: issue a certificate to the identity in the given token.
  \item $\alg{CA.Verify}(\pkca, \pkoidc, \cert, \id) \to \yes/no$: verify that the certificate matches the given identity, and that the certificate is valid for the given CA and IdP verification keys.
\end{itemize}

\noindent
This system is \emph{complete} if for all identities $\mathsf{id}$, we have:

\[
  \Pr\left[
  \begin{array}{l}
    \begin{array}{@{}ll@{}}
      (\skoidc, \pkoidc) & \gets \alg{OIDC.Generate}(\lambda),   \\
      (\skca, \pkca)     & \gets \alg{CA.Generate}(\lambda),     \\
      \tok               & \gets \alg{OIDC.Issue}(\skoidc, \id), \\
      \cert              & \gets \alg{CA.Issue}(\skca, \pkoidc, \tok) \\
    \end{array} \\
    \text{s.t.} \quad \alg{CA.Verify}(\pkca, \pkoidc, \cert, \id) = \yes
  \end{array}
  \right]
  = 1.
\]

\noindent
It is \emph{unforgeable} if, for all probabalistic polynomial-time adversaries $\adv$, we have:

\[
  \Pr\left[
  \begin{array}{l}
    \begin{array}{@{}ll@{}}
      (\skoidc, \pkoidc) & \gets \alg{OIDC.Generate}(\lambda), \\
      (\skca, \pkca)     & \gets \alg{CA.Generate}(\lambda), \\
      (\id, \cert)       & \gets \adv(\skca, \pkoidc)
    \end{array} \\
    \text{s.t.} \quad \alg{CA.Verify}(\pkca, \pkoidc, \cert, \id) = \yes
  \end{array}
  \right]
  \leq \negl(\lambda)
\]

\noindent
for some negligible function $\negl(\cdot)$. 
That is, it should be computationally infeasible for an adversary to produce a valid certificate for \emph{any} identity, even if they control the CA.

It is \emph{replay-protected} if, for all probablistic polynomial-time adversaries $\adv$, we have:

\[
  \Pr\left[
  \begin{array}{l}
    \begin{array}{@{}ll@{}}
      (\skoidc, \pkoidc) & \gets \alg{OIDC.Generate}(\lambda), \\
      (\skca, \pkca)     & \gets \alg{CA.Generate}(\lambda),  \\
      \tok               & \gets \alg{OIDC.Issue}(\skoidc, \id),  \\
      \cert              & \gets \alg{CA.Issue}(\skca, \pkoidc, \tok), \\
      \tok'              & \gets \adv(\pkca, \pkoidc, \id, \cert)
    \end{array} \\
    \text{s.t.} \quad \mathsf{OIDC.Verify}(\pkoidc, \tok', \id) = \mathtt{yes}
  \end{array}
  \right]
  \leq \mathsf{negl(\lambda)}.
\]

\noindent
That is, it should be computationally infeasible for an adversary seeing the certificate to create a valid OIDC token.  We argue that the construction in \cref{sec:design} meets these goals:

Completeness follows immediately from the construction.

Unforgeability follows from the proof-of-knowledge property of the ``proof of authentication'' implemented using the Guillou-Quisquater (GQ) identification scheme and the security of the RSA signature scheme.
Consider, towards contradiction, an adversary capable of forging certificates that a verifier accepts with non-negligible probability.
Because verifiers check the proof of knowledge of signature from the certificate on the token, this means that the adversary creates valid GQ signatures for a given public key.
These GQ signatures are a proof of knowledge for an RSA signature over the token.
This means that such an adversary could run the proof-of-knowledge extractor to create a valid signature over the OIDC token, violating the chosen-message security of the RSA scheme.

Replay-protection follows from the zero-knowledge property of the GQ scheme.
Suppose, towards contradiction, that some adversary can win the replay-protection game with non-negligible probability.
This means that they can produce tokens that pass OIDC verification.
Such verification involves checking a signature on the generated token.
This means that the adversary has been able to forge a token with a valid signature knowing only the public key of the CA, the public key of the OIDC identity provider, the identity, and the certificate.
We use this adversary to create an adversary $\adv'$ that simulates GQ signatures.
$\adv'$ receives $\pkoidc$ and creates a new $\pkca$ using $\alg{CA.Generate}$, and issues a certificate using a signed token requested from the OIDC signing oracle.
It can then pass these to $\adv$ and receives a valid, signed certificate.
Because these are distributed exactly as in the replay-protection game, $\adv'$ has exactly the same distribution as real GQ signatures, and any computationally bound algorithm will be able to tell these from honestly-created signatures with only negligible probability.

\end{document}